\documentclass[12pt]{article}
\usepackage{amsmath}
\usepackage{graphicx}
\usepackage{multirow}
\usepackage{citesort}

\newcommand{\spa}{\,}

\newcommand{\ep}{\epsilon}
\newcommand{\beq}{\begin{equation}}
\newcommand{\eeq}{\end{equation}}
\newcommand{\bea}{\begin{eqnarray}}
\newcommand{\eea}{\end{eqnarray}}
\newcommand{\re}[1]{(\ref{#1})}

\newcommand{\dimr}{d}

\newcommand{\physq}{P}
\newcommand{\rc}[1]{Z^{(#1)}}
\newcommand{\monintc}[2]{f^{(#1)}_{#2}(\ep)}
\newcommand{\monint}[2]{F^{(#1)}_{#2}(\ep)}
\newcommand{\numm}{N}
\newcommand{\mint}[2]{M^{(#1)}_{#2}(\ep)}
\newcommand{\mintc}[2]{a^{(#1)}_{#2}(\ep)}
\newcommand{\numfm}{N}
\newcommand{\fmint}[2]{\tilde{M}^{(#1)}_{#2}(\ep)}
\newcommand{\fmintc}[2]{c^{(#1)}_{#2}(\ep)}
\newcommand{\maxdiv}{{n_\text{max}}}
\newcommand{\divset}[1]{\{F_\text{div}^{(#1)}\}}
\newcommand{\intermm}[2]{\hat{M}^{(#1)}_{#2}(\ep)}

\newcommand{\cutoff}{\Lambda}

\newcommand{\diaA}{$\diaAeq$}
\newcommand{\diaB}{$\diaBeq$}
\newcommand{\diaC}{$\diaCeq$}
\newcommand{\diaD}{$\diaDeq$}
\newcommand{\diaE}{$\diaEeq$}
\newcommand{\diaF}{$\diaFeq$}
\newcommand{\diaG}{$\diaGeq$}
\newcommand{\diaH}{$\diaHeq$}
\newcommand{\diaI}{$\diaIeq$}

\newcommand{\diaAeq}{{T_{52}^f}}
\newcommand{\diaBeq}{{T_{54}^f}}
\newcommand{\diaCeq}{{T_{61}^f}}
\newcommand{\diaDeq}{{T_{62}^f}}
\newcommand{\diaEeq}{{T_{64}^f}}
\newcommand{\diaFeq}{{T_{71}^f}}
\newcommand{\diaGeq}{{T_{72}^f}}
\newcommand{\diaHeq}{{T_{81}^f}}
\newcommand{\diaIeq}{{T_{91}^f}}

\newcommand{\intmom}{p}
\newcommand{\cutdia}[2]{\Sigma^{\text{#1}}(#2)}
\newcommand{\cutdiafinite}[2]{\tilde{\Sigma}^{\text{#1}}(#2)}

\newcommand{\nlow}{{n_\text{low}}}
\newcommand{\nhigh}{{n_\text{high}}}

\newcommand{\paden}{n}
\newcommand{\padem}{m}
\newcommand{\pade}[4]{\Sigma^{\text{#1}}_{[#3/#4]}({#2}^2)}
\newcommand{\finalv}[3]{{\text{#1}}_{#2,#3}}

\newcommand{\highenergy}[2]{\Sigma^{\text{#1}}_{q \to \infty}({#2}^2)}

\newcommand{\PolyLog}{\text{Li}_4 \left( {\textstyle \frac 12} \right)}

\newcommand{\others}{\dots}

\begin{document}

\begin{titlepage}

{
\centerline{\normalsize\hfill  SFB/CPP-06-01}
\centerline{\normalsize\hfill  TTP06-02     }
\centerline{\normalsize\hfill hep-ph/0601165}
\baselineskip 11pt
{}
}

\vspace{0.5cm}
\begin{center}
  \begin{Large}
  \begin{bf}
$\boldsymbol{\ep}$-Finite Basis of Master Integrals for the
Integration-By-Parts Method 
  \end{bf}
  \end{Large}

  \vspace{0.8cm}

  \begin{large}
    K.G. Chetyrkin\footnote{On leave from Institute for
      Nuclear Research of the Russian Academy of Sciences, Moscow, 117312,
      Russia.}$^{,a}$, M.~Faisst$^a$, 
      C.~Sturm$^b$ {\normalsize and } M.~Tentyukov\footnote{On leave
      from BLTP, Joint Institute
             for Nuclear Research, Dubna, Russia.}${^{,a}}$
  \end{large}
  \vskip .7cm

         {\small $^a$ {\em
             Institut f\"ur Theoretische Teilchenphysik,
             Universit\"at Karlsruhe,
             D-76128 Karlsruhe, Germany}}

{\small $^b$ {\em  Dipartimento di Fisica Teorica, Universit{\`a} di
    Torino, Italy\\   INFN, Sezione di Torino, Italy}}

        \vspace{0.8cm}
{\bf Abstract}
\end{center}
\begin{quotation}
\noindent
It is shown that for every problem within dimensional regularization,
using the Integration-By-Parts method, one
is able to construct a set of master integrals  
such that each corresponding coefficient
function is {\em finite} in the limit of dimension equal to four. We
argue that the use of such a basis simplifies and stabilizes the
numerical evaluation of the master integrals. As an example we
explicitly construct the $\ep$-finite basis for the set of all QED-like
four-loop massive tadpoles. Using a semi-numerical approach based on
Pad\'{e} approximations we evaluate analytically the divergent and
numerically the finite part of this set of master integrals.

The calculations confirm the recent results of Schr\"oder and
Vuorinen. All the contributions found there by fitting the high
precision numerical results have been confirmed by direct analytical
calculation without using any numerical input.  
\end{quotation}
\end{titlepage}

\section{Introduction}

The evaluation of higher order perturbative corrections plays an
important role in testing the Standard Model on a high precision level
as well as in the search for new physics beyond the established
theory.  Especially given the increasing resolution of modern
experiments, it is mandatory to push ahead the efforts on the theory
side. Many reviews, see
e.g. \cite{Steinhauser:2002rq,Harlander:1998dq}, deal with the
application of such calculations. The present work focuses on the
actual evaluation of the integrals appearing when considering higher
order corrections.

Let us begin by noting that basically every analytical method
applicable for higher order calculations can effectively be put into
action only within the  ``calculation friendly''  environment provided by
dimensional regularization
\cite{Cicuta:1972jf,Ashmore:1972uj,'tHooft:1972fi}. Of course, computing
e.g. the 1-loop $\beta$ function one could use a variety of
approaches. However already at 2-loop level it is almost impossible to
avoid the use of dimensional regularization unless one is considering a
simple scalar toy model.

It should also be stressed that from a purely calculational point of
view the most valuable feature of dimensionally regularized integrals is
their similarity with the usual absolutely convergent integrals. As a
result almost all standard operations are mathematically justified.

In particular, the traditional Integration By Parts (IBP) method is based 
on  the possibility to discard the surface terms within this
framework\footnote{The possibility is by no means obvious for a general
  case of integrals containing {\em both} UV and IR divergences and
  requires a { dedicated} proof. The latter, in turn, is not possible
  without a proper { definition } of the very dimensional regularization
  for such integrals. To the best of our knowledge  the elaborated
  mathematically meaningful definition of dimensional regularization and
  its  main properties was presented in
  \cite{Chetyrkin:1983wh,book:Smirnov:1991jn} and reviewed in
  \cite{book:Smirnov:2002pj}.}
and   allows for generating exact identities
between different integrals. In a next step the resulting equations are
used to reduce every Feynman diagram to a limited set of mutually
independent so-called {\em master integrals}. 

This reduction procedure is a highly non-trivial problem by itself as
the growing number and complexity of IBP identities defies any attempt
of a straightforward solution starting from some number of loops and
external legs. Therefore the classical way of ``manually'' examining
the IBP identities seems to be unsuitable for such problems. A
prominent exception is the crack-down of the long-standing problem of
computing the generic moments of the structure functions in
deep-inelastic scattering at three-loop level
\cite{Vogt:2004mw} which has been completed  only recently. At the
moment there are a few approaches trying to automatize the procedure of
reduction to master integrals
\cite{Laporta:1996mq,Laporta:2001dd,Baikov:1996cd,Baikov:tadpoles:96,%
Baikov:2003zq,Tarasov:1998nx,Tarasov:2004ks,Schroder:2002re,%
Binoth:2003ak,Smirnov:2003kc,Smirnov:2005ky,Anastasiou:2005cb,%
Czakon:2005rk}. Once the reduction has been achieved one is left with
the evaluation of the resulting master integrals. A complete up-to-date
review about various methods of  calculating Feynman diagrams can be
found in \cite{book:Smirnov:2004ym}.

As is  well-known within the method of dimensional regularization, the
usual UV divergences manifest themselves as poles in $\ep = (4-d)/2$,
with $d$ being the  spacetime dimension. The calculation of the master
integrals is complicated by appearance of unphysical, so-called
spurious, poles in the coefficient functions which are often induced
during the reduction procedure. Through the appearance of these spurious
poles one is forced to compute extra terms in the  $\ep$-expansion of
the master integrals. In general the evaluation of every next term in
the $\ep$-expansion is  challenging, especially if the application of
numerical methods is necessary (see,
e.g.
\cite{Davydychev:2000na,Laporta:2002pg,Schroder:2005va,Czakon:2004wu,%
Czakon:2004wu:extra,Kalmykov:2005hb,Kniehl:2005yc}).

On the other hand, the choice of master integrals is 
by no means unique.

In the present work we discuss a convenient way of finding a new
basis of master integrals. In particular we show in the next two
sections that for every problem reducible with the help of the IBP
method to a set of master integrals the latter can be always chosen in
such a form that {\em all} the corresponding coefficient functions are
{\em finite} in the limit $\ep \to 0$. We call this basis the
$\ep$-finite basis. In section \ref{sec:vacpol.basis} we explicitly
construct such an $\ep$-finite basis for the set of all QED-like
four-loop massive tadpoles appearing in the first moment of the low
energy expansion of the vacuum polarization in QCD. Using a
semi-numerical approach based on Pad\'{e} approximations we evaluate
numerically all the masters of this $\ep$-finite set and confirm the
recent results of Schr\"oder and Vuorinen \cite{Schroder:2005va}. Then
we focus on analytical results which can be easily derived following
this method.  Our general comments are presented in section
\ref{sec:comm}.

\section{Generalities and Notation}

Let us consider a quantity $\physq$  which we want to compute up to 
$L$-loop level. The sum of all bare contributions $\physq_B$ is given by
\beq
\physq_B = \sum_{l=0}^L h^l \physq_B^{(l)}.
\label{PB}
\eeq 
Here we have introduced a formal parameter $h$ counting the number of
loops. The renormalized $L-$loop contribution $h^L \, \physq_R^{(L)}$ is
expressible through 
\beq \label{PR^L}
h^L \, \physq_R^{(L)} =  h^L \, \physq_B^{(L)} + \sum_{l=1}^L h^l \rc{l}
\, h^{L-l}\physq_B^{(L-l)}
\eeq
as a linear combination of the bare quantities $\physq_B^{(l)}$ and the
corresponding renormalization constants, here summarized in $h^l \,
\rc{l}$ for the $l$-loop contribution. We assume that the calculation of
the integrals up to $h^{L-1}$ order is well under control. This includes
the sum in the second term of equation \re{PR^L}. Therefore we
concentrate on the bare $L$-loop contribution $\physq_B^{(L)}$.

After the evaluation of all the traces in $\dimr$ dimensions each
$l$-loop contribution can be written as a linear combination of
$\dimr$-dimensional integrals $\monint{l}{i}$ and associated coefficient
functions $\monintc{l}{i}$
\beq
\physq_B^{(l)} = \sum_{i} \monintc{l}{i} \monint{l}{i}.
\label{eq:monomial}
\eeq
Here possible divergencies in $\ep$ can only result from the evaluation
of the various integrals $\monint{l}{i}$ whereas the functions
$\monintc{l}{i}$ remain finite in limit $\ep \rightarrow 0$. The
resulting number of integrals $\{\monint{l}{i}\}$ is in general
extremely high and therefore reduced by use of IBP relations to a small
set of $\numm$ master integrals $\{\mint{l}{i}\}$ of $l$-loop order. In
this framework the contribution $\physq_B^{(L)}$ can be written as
\beq
\label{eq:master} 
\physq_B^{(l)} = \sum_{i=1}^\numm \mintc{l}{i} \mint{l}{i}.
\eeq
Through the dependence on the spacetime dimension $\dimr$ of the IBP
relations the coefficient functions $\mintc{l}{i}$ in equation
\re{eq:master} can contain divergencies for $\ep \rightarrow 0$. In this
case the expansion in $\ep$ of the corresponding  master integrals
$\mint{l}{i}$ has to be evaluated deeper than to the finite contribution
in order to compensate for spurious poles. In the next section we will
construct a different set of $\numfm$ master integrals
$\{\fmint{l}{i}\}$ such that the expansion
\beq \label{eq:fmaster}
\physq_B^{(l)} = \sum_{i=1}^\numfm \fmintc{l}{i} \fmint{l}{i}
\eeq
leads to coefficient functions $\fmintc{l}{i}$ being finite in the limit
$\ep \rightarrow 0$. This new set of master integrals is called
$\ep$-finite basis.

\section{Construction of the $\boldsymbol{\ep}$-finite
  set\label{sec:constr}}

The comparison of equations \re{eq:monomial} and \re{eq:master}
\beq
\sum_{i} \monintc{l}{i} \monint{l}{i} = \sum_{i=1}^\numm \mintc{l}{i}
\mint{l}{i}
\eeq
suggests that the $\ep$-finite basis can be found in the set of
integrals $\{\monint{l}{i}\}$ as the coefficients $\monintc{l}{i}$
already expose the desired behavior. By definition 
all the initial integrals can be expressed in terms of master integrals
\beq \label{eq:mon.int}
\monint{l}{i} = \sum_{j=1}^{\numm} \mintc{l}{ij} \mint{l}{j}.
\eeq
Among this in general huge set of equations only the subset $\divset{l}$
exhibiting a divergent coefficient function $\mintc{l}{ij}$ is
interesting. Let $\mintc{l}{mn}$ be one of the functions showing the 
$1/\ep^\maxdiv$-divergency with the highest power $\maxdiv$. Starting
from the relation for $\monint{l}{m}$ involving $\mintc{l}{mn}$ we are
able to construct a new basis of master integrals by solving equation
\re{eq:mon.int} for the master integral associated with  $\mintc{l}{mn}$
\beq \label{eq:replace}
\mint{l}{n} = \frac{1}{\mintc{l}{mn}} \left( \monint{l}{m} -
  \sum_{\substack{j=1 \\ j\neq n}}^{\numm} \mintc{l}{mj} \mint{l}{j}
\right).
\eeq
As $\mintc{l}{mn}$ was divergent for $\ep \rightarrow 0$ the reciprocal
function is finite. By means of this replacement $\monint{l}{m}$ is now
a member of the new set of master integrals  $\{\intermm{l}{i}\}$
replacing $\mint{l}{n}$ in the following way:
\beq
\{\intermm{l}{i}\} = \{ \mint{l}{1}, \dots, \mint{l}{n-1}, \monint{l}{m},
\mint{l}{n+1}, \dots, \mint{l}{\numm}\}.
\eeq
In a next step the replacement \re{eq:replace} is performed for all
equations in the set $\divset{l}$.

This procedure is iterated until all coefficient functions in
$\divset{l}$ are finite. The thus obtained final set of master integrals
$\{\intermm{l}{i}\}$ fulfills the relation
\beq
\physq_B^{(l)} = \sum_{i} \monintc{l}{i} \monint{l}{i} = \sum_{i,j}
\monintc{l}{i} \mintc{l}{ij} \intermm{l}{j},
\eeq
where each appearing coefficient function is by construction finite. If
we now identify
\begin{gather}
\fmintc{l}{j} = \sum_{i} \monintc{l}{i} \mintc{l}{ij},\\
\fmint{l}{j} = \intermm{l}{j},
\end{gather}
we arrive at the desired form \re{eq:fmaster} and call the set of
integrals $\{\fmint{l}{i}\}$ the $\ep$-finite basis. We again stress the
fact that all the integrals in this $\ep$-finite basis need to be
evaluated only up to the finite part in their $\ep$-expansion, as by
construction no spurious poles arise.

\section{Applications to four-loop tadpoles\label{sec:vacpol.basis}}

\begin{figure}[t]
  \centering
  \includegraphics[height=2.1cm]{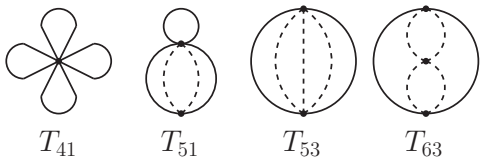}
  \caption{\label{fig:simple}Subset of master integrals fully
    analytically known and therefore not considered in the construction
    of the $\ep$-finite basis.}
\end{figure}

\begin{figure}[t]
  \centering
  \includegraphics[width=14.5cm]{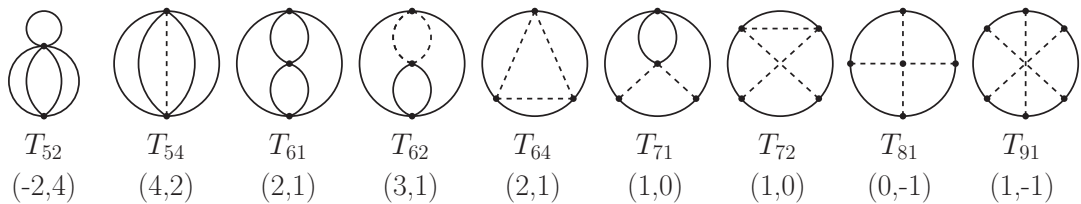}
  \caption{\label{fig:master}Subset of master integrals for which the
  $\ep$-finite basis is constructed. The two numbers in brackets
  $(n_1,n_2)$ are decoded as  follows: $n_1$ is the maximal power of a
  spurious pole $(\frac{1}{\ep})^{n_1}$ which could appear in front of
  the diagrams pictured above; ${n_2}$ is the maximal analytically
  known power $\ep^{n_2}$ of the   $\ep$-expansion of the  same integral
  as determined in \cite{Broadhurst:1992fi,Broadhurst:1996az} 
  (for $T_{52}$) and in \cite{Schroder:2005va,Schroder:2005db}
  for all the rest.
}
\end{figure}

The concept of the $\ep$-finite basis is now applied to the set of
master integrals appearing in the calculation of QED-like four-loop
tadpole diagrams. We concentrate on the subset of master integrals
shown in figures \ref{fig:simple} and \ref{fig:master} needed for the
calculation of the low
energy expansion of the four-loop vacuum polarization function,
completing the work in \cite{Chetyrkin:2004fq}. We denote  each of the
master integrals according to the following rule: after a capital letter
``T'' we write the number of lines in the given diagram. The second
number enumerates the different topologies exhibiting the same number of
lines. The diagrams of the first subset shown in figure \ref{fig:simple}
are  analytically fully known and expressible in terms of
$\Gamma$-functions. We thus  construct the $\ep$-finite basis only  for
the second subset of the  most complicated  master integrals shown in
figure \ref{fig:master}. This is done by treating the coefficient
functions of the ``simple'' diagrams in the set of equations
$\divset{4}$ as finite.

\subsection{Construction of the $\boldsymbol{\ep}$-finite
  set\label{sec:constr.basis}}

\begin{figure}[t]
  \centering
  \includegraphics[height=4.4cm]{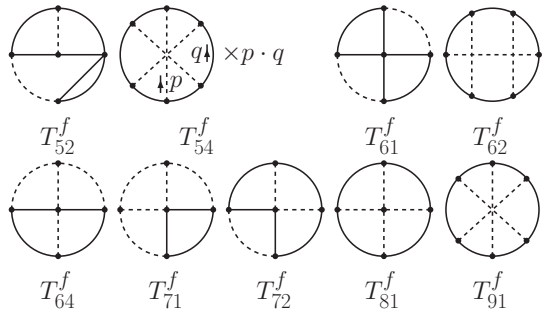}
  \caption{\label{fig:epfb}Diagrams of the $\ep$-finite basis.}
\end{figure}

Following the description in section \ref{sec:constr} we find the
diagrams shown in figure \ref{fig:epfb} as the $\ep$-finite
basis. Here the name of the $\ep$-finite master integrals is deduced
by using the name of the original master, which had been replaced, and
the additional superscript letter ``f''. Note that except for $T_{54}^f$
all members of the $\ep$-finite basis pictured in figure \ref{fig:epfb}
are just scalar Feynman integrals. The integrand corresponding to
$T_{54}^f$ contains a scalar product of two line momenta in its
numerator.
According to equation~(\ref{eq:mon.int}) the integrals of the
$\ep$-finite basis can be expressed in terms of the original master
integrals of figure~\ref{fig:simple} and \ref{fig:master}. This has been
achieved by using the traditional IBP method in combination with the
Laporta Algorithm. In this context the IBP identities have been
generated with numerical values for the powers of the propagators and
the irreducible scalar products. In the next step, the resulting system
of linear equations has been solved, by expressing systematically
complicated integrals in terms of simpler ones and substituting the
solutions into all the other equations. For this reduction to the
original master integrals an automated program, which is based on
{\tt{FORM3}} \cite{Vermaseren:2000nd,Vermaseren:2002rp} has been
developed in which also partially ideas described in
ref. \cite{Laporta:2001dd,Mastrolia:2000va,Schroder:2002re,%
Anastasiou:2004vj} have been implemented. The coefficient functions of
the integrals in the linear system of equations depend on the spacetime
dimension $d$ and need to be simplified during the solving
procedure. This has been done with the program {\tt{FERMAT}}
\cite{Lewis}. Having expressed the integrals of figure~\ref{fig:epfb} in
terms of the original basis and following than
equation~(\ref{eq:replace}), one obtains the relations which replace the
original master integrals by the $\ep$-finite ones. As these equations
are rather lengthy, they are summarized in
appendix~\ref{app:OldFromNew}.

\subsection{Calculation of the $\boldsymbol{\ep}$-finite
  basis\label{sec:pade}}

In order to calculate the results for the $\ep$-finite basis we follow
the method based on the ideas in ref. \cite{Fleischer:1994ef} and
already described in \cite{Faisst:2004kz}. One line of the four-loop
tadpole under consideration is cut, thus obtaining a three-loop
self-energy diagram. Since there is no other four-momentum involved, the
self-energy diagram only depends on the momentum flowing through the
applied cut. Therefore we are able to immediately perform the angular
part of the remaining loop integration. Following this procedure, the
calculation of the four-loop tadpole is transformed to the problem of
evaluating a three-loop self-energy and a one-dimensional numerical
integration over the modulus of the momentum through the cut.

The resulting three-loop diagrams are treated with the help of
asymptotic expansions. Using the program packages {\tt EXP}
\cite{Seidensticker:1999bb,Harlander:1997zb}, {\tt MATAD}
\cite{Steinhauser:2000ry} and {\tt MINCER} \cite{Larin:1991fz} one is
able to compute the low and high energy expansion of the self-energy
diagram. In a final step we combine the obtained high and low energy
expansions in a Pad\'e approximation as described in
\cite{Chetyrkin:1996cf}. The resulting function is integrated
numerically up to a cutoff $\cutoff$ as the integration up to infinity
is in general diverging. The remaining integration region from $\cutoff$
to infinity can be treated by direct analytical integration of the high
energy expansion. Therefore one is able to compute the structure of the
$\ep$-poles analytically. We checked that the result for  the finite
part of the master integrals is cutoff-independent. In addition we
tested that for finite tadpole diagrams like \diaI{} the difference
between the introduction of $\cutoff$ and the numerical integration up
to infinity is below the error of the numerical integration. A more
detailed description of the procedure can be found in
\cite{Faisst:diss}. For diagram \diaA{} we explicitly show the steps in
the next subsection. 

Note that in general we have a vast freedom in choosing the members of
the $\ep$-finite basis. Thus we give some additional remarks about
selecting appropriate diagrams. For the applicability of the
Pad\'e method in the described way the self-energy diagram resulting
after cutting one line should not contain any massless cuts (see
\cite{Faisst:diss}). Nevertheless, for each integral shown in figure
\ref{fig:epfb} at least two possible cuts leading to two different
self-energy diagrams could easily be found. This also provides us with a
additional possibility to check the results.

The self-energy integrals appearing after cutting one line might contain
UV and/or IR divergences. In general this is not a problem for the
presented method as one is always able to subtract suitable diagrams 
containing the same sub-divergencies but of lower loop order, thus
ending up with a self-energy function finite in the limit $\ep
\rightarrow 0$.  In appendix \ref{app:diaB} we demonstrate this
procedure explicitly for one diagram. However, a simple power counting
applied to the diagrams of figure \ref{fig:epfb} demonstrates that
excluding diagram $T_{52}^f$ every possible cut leads to {\em finite}
self-energies. In fact, the minimal presence of any divergencies was one
of the main principles used by us in choosing suitable candidates for
the $\ep$-finite basis.

\subsection{Treatment of the basis integral $\boldsymbol{\diaAeq}$
  \label{subsect:T52}} 

This diagram has two possible equivalent cuts which lead to the same
finite self-energy function $\cutdia{\diaA}{\intmom}$ related to the
diagram 
\begin{equation}
  \label{eq:self.diaA}
  \cutdia{\diaA}{\intmom} =
  \parbox{4cm}{\includegraphics[width=4cm]{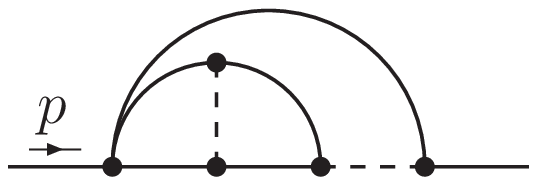}}
\end{equation}
resulting from the cut procedure. In general one might be left with a
diagram which does not allow for a cut resulting in a finite self-energy
function. In appendix \ref{app:diaB} we present an alternative
proceeding for diagram \diaA{} involving a cut leading to a divergent
self-energy function. As outlined in the previous section we
calculate the high and low energy expansion using {\tt EXP}, {\tt MATAD}
and {\tt MINCER} before combining these results in a
Pad\'e approximation
\begin{equation}
  \label{eq:pade}
f_{[{\paden}/{\padem}]}(z) = \frac{a_0 + a_1 z + \dots + a_{\paden}
   z^{\paden}}{b_1 z + \dots + b_{\padem} z^{\padem}}.
\end{equation}
Given $\nlow$ terms of the low energy and $\nhigh$ terms of the high
energy expansion we are able to evaluate a Pad\'e approximant
$\pade{\diaA}{\intmom}{\paden}{\padem}$ of the self-energy function in
equation \re{eq:self.diaA} for all values of $n$ and $m$  obeying the
relation
\begin{equation}
  \label{eq:padepower}
  \nlow + \nhigh = \paden + \padem.
\end{equation}
This function we are then able to integrate numerically up to a cutoff
$\cutoff$ according to
\begin{align}
  \diaAeq_{[{\paden}/{\padem}]}(\cutoff) &= \int_{\cutoff}
  \frac {d^{\dimr}\intmom}{(2\pi)^{\dimr}} \, \frac 1{\intmom^2 +
    m_{\text{cut}}^2} \pade{\diaA}{\intmom}{\paden}{\padem}\\
&=\frac{2\pi^{\dimr/2}}{(2\pi)^{\dimr}} \int_0^{\cutoff} d \intmom \,
\frac{{\intmom}^{\dimr-1}}{\intmom^2 +
    m_{\text{cut}}^2} \pade{\diaA}{\intmom}{\paden}{\padem}
  \label{eq:diagramA}
\end{align}
where $m_{\text{cut}}=m$ if we cut a massive and $m_{\text{cut}}=0$ if
we cut a massless line.

The remaining integration from the cutoff $\cutoff$ to infinity is
performed by explicit analytical integration over the high energy
expansion $\highenergy{\diaA}{\intmom}$ of the self-energy
\re{eq:self.diaA}. In contrast to \re{eq:diagramA} here one just has to
replace the Pad\'e approximant by  the high energy expansion
\begin{equation}
  \label{eq:analytic}
  \diaAeq_{,\, q \to \infty}(\cutoff) =
  \frac{2\pi^{\dimr/2}}{(2\pi)^{\dimr}} \int_{\cutoff}^{\infty} d
  \intmom \, \frac{{\intmom}^{\dimr-1}}{\intmom^2 + m_{\text{cut}}^2}
  \highenergy{\diaA}{\intmom}.
\end{equation}
For $m_{\text{cut}}=0$ or if we expand the denominator of the integrand
in case of $m_{\text{cut}}=m$ this integral has the simple form
\begin{equation}
  \label{eq:analyticpart}
  \diaAeq_{,\, q \to \infty}(\cutoff) =
\sum_{i=0}^{i_\text{max}} \int_{\cutoff}^{\infty} d \intmom \, a_i(\ep) \,
\intmom^{- n_i(\ep)}
\end{equation}
with noninteger positive exponents $n_i(\ep)$, at most linearly  dependent
on $\ep$, and   can be evaluated analytically 
\begin{equation}
  \label{eq:analyticres}
  \diaAeq_{,\, q \to \infty}(\cutoff) =
\sum_{i=0}^{i_\text{max}} \frac{a_i(\ep)}{- n_i(\ep)+1} \, (\cutoff)^{-
  n_i(\ep)+1}.
\end{equation}
As some of the exponents $n_i(\ep)$ are linear dependent on $\ep$, one
has to evaluate the high-energy expansion
$\highenergy{\diaA}{\intmom}$ up to the order linear in $\ep$. Following
this way the divergent part can be fully recovered through this
analytical integration as we have chosen the diagrams and the cut in
such a way that the self-energy is finite for $\ep \to 0$. The final
result for the diagram \diaA{} is then given by the sum
\begin{equation}
  \label{eq:finaldiaA}
\diaAeq_{[{\paden}/{\padem}]} = 
  \diaAeq_{[{\paden}/{\padem}]}(\cutoff) + \diaAeq_{,\, q \to
    \infty}(\cutoff)
\end{equation}
of the two contributions \re{eq:diagramA} and \re{eq:analytic}. Here
numerical evaluations only enter in the calculation of  the first term
resulting from the Pad\'e approximation. Therefore we obtain the full
pole part of the diagrams analytically. In addition, we stress again,
that the evaluation of any higher term in the $\ep$-expansion is not
needed.

\begin{table}[t]
  \centering
\begin{tabular}{|lc|c|c|c|} \hline
\multicolumn{2}{|c|}{$\finalv{\diaA}{\nlow}{\nhigh}$}
&\multicolumn{3}{c|}{$\nhigh$} \\
\multicolumn{2}{|c|}{} & 7 & 8 & 9\\ \hline
\multirow{4}{1.3cm}{$\nlow$}
& 4 & $27.30068104(1290)$ & $27.30068323(70)\phantom{00}$ &
$27.30068178(198)$ \\ \cline{2-5}
& 5 & $27.30067759(59)\phantom{00}$ & $27.30067456(1180)$ &
$27.30068017(86)\phantom{0}$ \\ \cline{2-5}
& 6 & $27.30068220(998)\phantom{0}$ & $27.30068156(146)\phantom{0}$ &
$27.30068059(36)\phantom{0}$ \\ \cline{2-5}
& 7 & $27.30067826(492)\phantom{0}$ & $27.30068069(34)\phantom{00}$ &
$27.30068067(5)\phantom{00}$\\ \hline
\end{tabular}
  \caption{\label{tab:diaA}Results for diagram \diaA{} depending on
    different number of low and high energy input terms. In brackets we
    give the error of the procedure deduced from the standard deviation.}
\end{table}
According to equation \re{eq:padepower} we are left with some freedom
how to choose the Pad\'e parameters $\paden$ and $\padem$ for a given
value of $\nlow$ and $\nhigh$. Thus we average for each combination of
$\nlow$ and $\nhigh$ the results of all possible $n$,
$m$-values. The average value $\finalv{\diaA}{\nlow}{\nhigh}$ is
what we give as the final numerical result for a given set of high and
low energy terms. The resulting standard deviation is a measure of  the
accuracy of this numerical approach. The values obtained for diagram
\diaA{} depending on different values of $\nlow$ and $\nhigh$ are given
in table \ref{tab:diaA}. The result is nicely converging for increasing 
number of input terms.

\subsection{Results for the $\boldsymbol{\ep}$-finite basis}

Following this method we evaluated all the diagrams of the $\ep$-finite
basis shown in figure \ref{fig:epfb} up to the finite part. The
numerical value of the constant term is obtained by using again nine
terms in the high and seven terms in the low energy expansion of the
self energy function arising after the cut. For each integration we use
for the euclidian integrals in analogy to \cite{Chetyrkin:2004fq}
\begin{equation}
  \frac{\mu^{4-d}}{N} \int \frac {d^d l}{(2\pi)^d} \dots
\end{equation}
with the normalization factor
\begin{equation}
    N=\ep\*\mu^{4-d}\int   \frac {d^d l}{(2\pi)^d}\frac{1}{(l^2 + m^2)^2}
    =\frac{1}{16\*\pi^2}\*\Gamma(1+\ep)\*
    \left(\frac{m^{2}}{4\*\pi\mu^2}\right)^{-\ep}.
\end{equation}
In the following we consider $m=1$. The results for the $\ep$-finite
basis then read:
\begin{align}
\diaAeq = & \,
\frac{3\,\zeta_3}{2\,{\ep}^2}
+ \frac{15\,\zeta_3}{2\,\ep}
- \frac{91\,{\pi}^4}{360\,\ep}
- \frac{4\,{\pi }^2\,\log^2 2}{3\,\ep}
+ \frac{4\,{\log^4 2}}{3\,\ep} \nonumber \\
&+ \frac{32\,\PolyLog}{\ep}
+ 27.30068067(5),\label{tf52} \\
\diaBeq = & \, \frac{5\,\zeta_5}{4\,\ep} - 6.762240238547(2),
\label{tf54}
\\
\diaCeq = & \, \frac{5\,\zeta_5}{\ep}-29.703462427815(3),
\label{tf61}
\\
\diaDeq = & \, 2.440345823350757(6),
\label{tf62}
\\ \displaybreak[0]
\diaEeq = & \, \frac{5\,\zeta_5}{\ep}-18.026245978729184(4),
\label{t64}
\\\displaybreak[0]
\diaFeq = & \, \frac{5\,\zeta_5}{\ep}-5.19831391(6),
\label{tf71}
\\\displaybreak[0]
\diaGeq = & \, \frac{5\,\zeta_5}{\ep}-26.4964794044474738(5),
\label{tf72}
\\
\diaHeq = & \, \frac{5\,\zeta_5}{\ep}-22.53760348(3),
\label{tf81}
\\
\label{tf91}
\diaIeq = & \, 1.808879546207(2).
\end{align}
We emphasize that all divergent parts of these integrals are obtained
analytically without any numerical input and that higher orders in the
$\ep$-expansion are not needed. A further investigation on the analytic
results presented here is given in the next section. By applying the
relations from appendix \ref{app:OldFromNew} the results for the
$\ep$-finite basis can be related to the master integrals in
figure \ref{fig:master}. We find full agreement with the values found in
\cite{Schroder:2005va} as it is shown in table \ref{tab:compare}.
\begin{table}[t]
  \centering
  \begin{tabular}{c|c|c}
diagram & Pad\'e method & \cite{Schroder:2005va}\\ \hline
\diaA & 27.30068067(5) & 27.30068068 \\ \hline
\diaB & -6.762240238547(2) & -6.762240238546 \\ \hline 
\diaC & -29.703462427815(3) & -29.703462427816 \\ \hline
\diaD & 2.440345823350757(6) & 2.440345823350754 \\ \hline
\diaE & -18.026245978729184(4) & -18.026245978729184 \\ \hline
\diaF & -5.19831391(6) & -5.19831389 \\ \hline
\diaG & -26.4964794044474738(5) & -26.4964794044474733 \\ \hline
\diaH & -22.53760348(3) & -22.53760347 \\ \hline
\diaI & 1.808879546207(2) &   \phantom{-1}1.808879546208 \\ \hline
  \end{tabular}
  \caption{\label{tab:compare}Comparison of the constant part of the
    $\ep$-finite basis with results extracted from
    \cite{Schroder:2005va}. Note that in the last column we only keep
    the accuracy needed for this comparison, whereas in
    \cite{Schroder:2005va} higher precision results are available.}
\end{table}

\subsection{ Analytical results for the master
  integrals\label{analytic_masters}} 

In subsection \ref{subsect:T52} we discussed a method which can be
used to find the pole part of any massive four-loop tadpole
analytically. More generally, this procedure is formulated in the
following statement: the pole part of any $(L+1)$-loop tadpole is 
obtained  analytically  in terms of pole and  finite parts of properly
constructed $L$-loop tadpoles and $L$-loop massless integrals. An
explicit construction of such an algorithm for the particular case of
three-loop completely massive tadpoles can be found in
\cite{Chetyrkin:1998fm}.

It is instructive to investigate in some detail the internal mechanism
of the cancellation of the spurious poles after the ``old'' basis set is
expanded in terms of the new $\ep$-finite basis. As we will see, the
very existence of such cancellations leads to additional  analytical
results for master integrals. 

With $s_n$ we denote the maximal power of spurious poles which could
appear in the coefficient function of an ``old'' master integral
$T_n$. The values of $s_n$ for all non-trivial masters is given in
figure \ref{fig:master}. Let us start with integral $T_{71}$. As we find
a value of $s_{71} = 1$ one expects that all coefficient functions
should be at least proportional to $\ep$ when expressing $T_{71}$ in
terms of $\ep$-finite basis. Here we, of course, exclude terms
proportional to the analytically known integrals $T_{41}, \,T_{51},\,
T_{53}$ and $T_{63}$ which had by construction been excluded from
the $\ep$-finite basis.  The  explicit result  reads
\beq T_{71} = \ep\,(4 \, T_{64}^f + 6 \,
T_{71}^f + 2 \, T_{72}^f) + O(\ep^2)\, T^f + \others , 
\label{T71:a}
\eeq 
in full agreement with our expectations. Here we denote by $O(\ep^2)\,
T^f$ terms with coefficient functions suppressed by at least as $\ep^2$
at $\ep \to 0$ and $\others$ represents  contributions
coming from analytically known master integrals. Equation~\re{T71} also 
carries important information about the {\em finite} part of
$T_{71}$. Indeed, as the pole parts of the new master integrals are
known analytically, one could easily use equation\re{T71} to get  again 
analytically the pole and the finite part of $T_{71}$ leading to the
result
\beq
T_{71} = - \frac{1}{6 \ep^4} - \frac{3}{2 \ep^3} - \frac{26}{3
  \ep^2} - \frac{\zeta_3}{\ep^2} - \frac{41}{\ep} - \frac{10 \zeta_3}{3
  \ep} - \frac{\pi^4}{60 \ep} - \frac{7\pi^4}{30} + 3 \zeta_3 -
\frac{1039}{6} + 53 \zeta_5  +O(\ep). 
\label{T71}
\eeq
Repeating the same procedure for the integrals $T_{52},T_{54}, T_{61},
T_{62}, T_{64}$ and $T_{72}$ we obtain the following results:
\begin{align}
T_{52} =& -8\ep^2\, T_{52}^f + O(\ep^3)\, T^f +\others \label{T52} \\
=& - \frac{2}{{\ep}^4} - \frac{29}{3\,{\ep}^3}
- \frac{163}{6\,{\ep}^2} - \frac{601}{12\,\ep} - \frac{635}{24} -
\frac{112\,\zeta_3}{3} + \ep \, \bigg( \frac{4549}{16} +
  \frac{136\,{\pi}^4}{45} \nonumber \\
& + \frac{32\,{\pi}^2\,\log^2 2}{3} -  \frac{32\,\log^4 2}{3} -
256\,\PolyLog - \frac{952\,\zeta_3}{3} \bigg) +O(\ep^2),
\end{align}
\begin{align}
T_{54} =&  \ep^3\, \frac{256}{3}\left\{ T_{52}^f + T_{62}^f  + \frac 13 \,
  T_{91}^f \right\} + O(\ep^4)\, T^f +\others \label{T54} \\
=& -\frac{1}{\ep^4} - \frac{9}{2\,{\ep}^3} -\frac{415}{36\,{\ep}^2} -
\frac{4991}{216\,\ep} - \frac{89383}{1296} + \ep \, \bigg(
-\frac{3367679}{7776} + \frac{1792\,\zeta_3}{9} \bigg) \nonumber \\
& + \ep^2 \, \bigg( - \frac{137735095}{46656} +
\frac{68992\,\zeta_3}{27} - \frac{4352\,{\pi }^4}{135} -
\frac{1024\,{\pi}^2\,\log^2 2}{9} \nonumber \\&+ \frac{1024\,\log^4
  2}{9} + \frac{8192\,\PolyLog}{3} \bigg) + O(\ep^3)
{},
\end{align}
\begin{align}
T_{61} =& \, \ep^2\,\left\{ 24\, T_{52}^f +6\, T_{61}^f  + 12\, T_{64}^f +
  18\, T_{71}^f +6\, T_{72}^f \right\} + O(\ep^3)\, T^f +\others
\label{T61} \\
=& \, \frac{3}{2\,{\ep}^4} +\frac{19}{2\,{\ep}^3}
+\frac{67}{2\,{\ep}^2} +\frac{127}{2\,\ep} - \frac{3\,\zeta_3}{\ep} -
\frac{237}{2} + 97\,\zeta_3 - \frac{{\pi }^4}{20} + \ep \, \bigg(
-\frac{3969}{2} \nonumber \\ 
& - \frac{559\,{\pi }^4}{60} -32\,{\pi }^2\,\log^2 2
+32\,\log^4 2 +768\,\PolyLog + 1261\,\zeta_3 \nonumber\\
& + 189\,\zeta_5 \bigg) +O(\ep^2),
\end{align}
\begin{align}
T_{62} =& \, \ep^2\,\left\{ 16\, T_{52}^f +16\, T_{62}^f  +
  \frac{32}{3}\, T_{91}^f \right\} + O(\ep^3)\, T^f +\others \label{T62}
\\
=& \, \frac{2}{3\,{\ep}^4} +\frac{4}{{\ep}^3} +\frac{38}{3\,{\ep}^2}
+\frac{44}{3\,\ep} + \frac{16\,\zeta_3}{3\,\ep} - 118 + 88\,\zeta_3
-\frac{4\,{\pi }^4}{15} + \ep \, \bigg( - 1156 \nonumber \\
&-\frac{374\,{\pi }^4}{45} -\frac{64\,{\pi }^2\,\log^2 2}{3}
+\frac{64\,\log^4 2}{3} + 512\,\PolyLog
+\frac{2152\,\zeta_3}{3}\nonumber \\
&+96\,\zeta_5 \bigg) +O(\ep^2),
\end{align}
\begin{align}
T_{64} =& \, 9 \ep^2\, T_{71}^f +  O(\ep^3)\, T^f +\others \label{T64}
\\
=& \, \frac{1}{4\,{\ep}^4} + \frac{3}{2\,{\ep}^3} + \frac{9}{2\,{\ep}^2}
+ \frac{2}{\ep} +\frac{13\,\zeta_3}{2\,\ep} - \frac{147}{2} +
30\,\zeta_3 - \frac{5\,{\pi}^4}{8} + \ep \, \bigg( - 624 \nonumber
\\ 
&- \frac{27\,{\pi}^4}{10} + 45\,\zeta_3 + \frac{693\,\zeta_5}{2} \bigg)
+O(\ep^2), 
\end{align}
\begin{align}
T_{72} =& \, \ep\,\left\{ 2 \, T_{64}^f + 3 \, T_{71}^f \right\} +
O(\ep^2)\, T^f +\others \label{T72} \\
=& - \frac{1}{6\,{\ep}^4} - \frac{3}{2\,{\ep}^3}-
\frac{26}{3\,{\ep}^2} - \frac{\zeta_3}{2\,{\ep}^2} - \frac{41}{\ep} +
\frac{\zeta_3}{6\,\ep} - \frac{{\pi}^4}{120\,\ep} - \frac{1039}{6} -
\frac{29\,{\pi }^4}{120} \nonumber \\
& + \frac{17\,\zeta_3}{2}  +\frac{43\,\zeta_5}{2} +O(\ep)
{}.
\end{align}
It is remarkable, that depth of the expansion in $\ep$ of these
analytical results for ``old'' masters 
$T_{54},  T_{61}, T_{62}, T_{64}, T_{71} $ and   $T_{72}$
listed above are in full
agreement with  the depth obtained in \cite{Schroder:2005va} by fitting
the high precision numerical results.

It is even more remarkable that  in mostly  all of  the cases  the
highest power ${\ep}^{d_n}$ of the analytically accessible part of the
expansion is correlated with the maximal power of the corresponding
spurious pole $s_n$ introduced before, such that
\beq
d_n = s_n - 1.
\label{di=si-1}
\eeq
Thus, we observe some rationale as for how many terms of the
$\ep$-expansion of a master integral can be computed analytically.

However, equation \re{di=si-1} seems to be violated in case of the
integrals $T_{54}$, $T_{62}$ and $T_{91}$. Here we observe the relation 
\beq
d_i = s_i - 2
{}.
\label{di=si-2}
\eeq
At first glance equation~\re{di=si-2} is in apparent contradiction to
the very concept of the $\ep$-finite basis. Indeed, consider, for
example, the integral $T_{54}$. Given the value $s_{54}=4$ from
figure~\ref{fig:master}, it is known that this integral sometimes
appears accompanied by a spurious pole $\frac{1}{\ep^4}$. A naive
application of equation \re{T54} tells us that after expanding $\frac
1{\ep^4} T_{54}$ in terms of the $\ep$-finite basis the integrals
$T_{52}^f$, $T_{62}^f$ and $T_{91}^f$ would have {\em non}-$\ep$-finite
coefficients! The problem is solved by noting that these $\frac
1{\ep^4}$ poles always appear in one and the same combination, namely
\beq
T = \frac{1}{128}\frac{1}{\ep^4}\left\{
9\, T_{54}  - 48\,\ep \, T_{62} + 256\, \ep^3\, T_{91}
\right\} \label{inique_combination}
{}.
\eeq
As expected the result of expanding $T$ in the $\ep$-finite basis has
finite coefficients in front of non-trivial masters:
\begin{align}
T =&\, 11\,T_{52}^f - 8\,T_{54}^f - 8\,T_{61}^f + 11\,T_{62}^f +
6\,T_{64}^f + 18\,T_{71}^f + 10\,T_{72}^f \nonumber\\
& - \frac{73}{3}\,T_{91}  + O(\ep)\, T^f +\dots {} \label{T}
\end{align}
Hence, this relation should lead to an extra analytical constraint on
$T_{52}$, $T_{62}$ and $T_{91}$. Indeed,
equations~(\ref{tf52}-\ref{tf91}) allows us  to compute analytically the
pole  part of the right hand side of \re{T} and, simultaneously, the
same pole part could expressed through  Taylor expansion the right hand
side of \re{inique_combination}  at  $\ep \to 0 $. As a  result we get 
\begin{multline} \label{extra_constraint}
\frac{1}{4} \, T_{91,0} + \frac{9}{1024}\,T_{54,3} -
\frac{3}{64}\,T_{62,2} = \frac{5511907345}{31850496} +
\frac{89\,{\pi}^4}{1440} + \frac{{\pi}^6}{56} \\ -
\frac{11\,{\pi}^2\,\log^2 2}{6}+ \frac{11\,\log^4 2}{6} + 44\,\PolyLog
- \frac{4103\,\zeta_3}{144} - {\zeta_3}^2 - \frac{273\,\zeta_5}{8},
\end{multline}
where we have used the notation:
\beq
T_{n} = \sum_{i=n_{\mathrm{min} }}^{\infty} \ep^i\, T_{n,i} 
\label{taylor}
{}.
\eeq
The integrals $T_{54}$ and  $T_{62}$, being of a ``sunset''-like type,
are  certainly much more suitable for a numeric evaluation than $T_{91}$
(for an exhaustive review see, e.g. \cite{Groote:2005ay}). Moreover, the
integral $T_{91}$ seems to be rather complicated for a numerical
treatment as illustrated by the fact that in \cite{Schroder:2005va} it
was computed  with significantly less precision than all others.  Thus,
it is natural to take \re{extra_constraint} as an equation for the
calculation of $T_{91,0}$. Substituting $T_{62,2}$ and $T_{54,3}$ as
found in \cite{Schroder:2005va} one obtains immediately without any
further calculation
\beq
T_{91,0} =1.808879546208334741426364595086952090
\label{T91_0_N}
\eeq
which should be compared to the original result given in
\cite{Schroder:2005va} (for further discussion see the next section)
\beq
T_{91,0} =1.80887954620833474
\label{T91_0_N_schreder}
{}.
\eeq

\section{General Comments\label{sec:comm}}

There is a significant amount of freedom in  the choice of specific
members of the $\ep$-finite basis.  As we have seen in subsection
\ref{sec:pade} this freedom can be used in order to greatly  simplify
the evaluation of the master integrals. In fact, the very idea of the
$\ep$-finite basis appeared after facing problems in trying to evaluate
the ``old'' master integrals  with the help of the Pad\'e  method.

An important feature of the $\ep$-finite basis is extremely simple
structure of the UV and IR divergences of its members. Indeed, except
for $T_{81}$ and $T_{91}$ all members of the ``old'' basis set do suffer
from overlapping UV divergences. Whereas all new masters but one are
either finite or possess only mild superficial UV divergences. The
integral $T_{52}^f$ is not primitively divergent but its only UV
divergent subgraph could be very easily isolated as demonstrated in
appendix \ref{app:diaB}.

The fact that the IBP identities sometimes provide some nontrivial
constraints on the {\em values} of master integrals is, of course, not
entirely new.  For example, in \cite{Czakon:2004bu} the divergent parts
of some of the irreducible four-loop completely massive tadpoles,
appearing in the calculation of the four-loop QCD $\beta$ function, have
been obtained from the finiteness of the same integral with higher
powers of denominators. However, we have already discussed in section
\ref{analytic_masters} that the evaluation of the pole part of a given
four-loop tadpole is not a problem  at all. We are only aware of few
cases when IBP identities were actually used to evaluate the finite term
or terms proportional to a nonnegative integer power of $\ep$ in the
$\ep$-expansion of nontrivial four-loop master integrals.

In \cite{Schroder:2005db} the order $\epsilon$ coefficient of the master
integral $T_{54}$ (there denoted as BB4) could be deduced, as it had
been the only analytically unknown term entering the cancellation of the
divergent contributions during the renormalization procedure when
calculating the so-called $\rho$ parameter. Furthermore, the coefficient
of order $\epsilon^2$ was subsequently obtained by combining the
numerically known value with the basis of transcendentals from
\cite{Mastrolia:thesis}. Within our approach {\em all} the terms up to
the $\ep^2$ contribution are directly accessible by the use of
$\ep$-finite basis without any need for either a high-precision
numerical evaluation or the basis of transcendentals which could
appear.

Schematically, the relations which were used to fix $T_{54,1}$ in
\cite{Schroder:2005db} can be written as
\beq
\rho_B - \mbox{(known UV sub-divergencies)} = {\cal O}(\ep^0)
\label{rho1}
\eeq
and
\beq
\rho_B - \mbox{(known UV sub-divergencies)} = \alpha_s^3\,T_{54,1}/\ep
+\mbox{known terms}
\label{rho2}
{}
\eeq
with known UV sub-divergencies. Equations \re{rho1} and \re{rho2}
directly leads to the value of $T_{54,1}$. Nevertheless relations like
(\ref{rho1}) involving the cancellation of divergencies can not,
obviously, provide any new information in addition to what is obtainable
from the use of the $\ep$-finite basis. Furthermore there is no
systematic way to find such relations, not to mention the requirement
that all renormalization constants have to be known analytically.

The fact that such relations can sometimes be found by chance is also
well illustrated in the next example. In section \ref{analytic_masters}
equation \re{extra_constraint} was derived in a regular way as natural
by-product of the properties of the constructed $\ep$-finite
basis. Recently a paper has appeared where the same relation
(\ref{extra_constraint}) has been independently obtained in a similar
way like the result for $T_{54,1}$ in \cite{Schroder:2005db}. The
authors of reference \cite{YS_MS} have computed the so-called decoupling
function $\zeta_g$ for the QCD quark gluon coupling constant in
four-loop approximation. The bare function $\zeta^0_g$ is essentially
composed of three pieces: 
\begin{equation}
\zeta_g^0=\frac{\tilde\zeta_1^0}{\tilde\zeta_3^0\sqrt{\zeta_3^0}}
\label{zetag0}
{},
\end{equation}
where the three auxiliary bare functions on the right side
$\tilde\zeta_1^0, \,\tilde\zeta_3^0$ and $ \zeta_3^0$ are related to the
heavy quark containing contributions to the ghost-ghost-gluon vertex,
the gluon propagator and the  ghost propagator respectively (for more
details see
\cite{Chetyrkin:1998un,Schroder:2005db,Chetyrkin:2005ia}). Once the
three bare functions are determined and the physically relevant quantity
$\zeta_g^0$ is constructed one can renormalize $\zeta_g^0$ directly with
the help of (long-available from
\cite{vanRitbergen:1997va,Chetyrkin:1997dh,Vermaseren:1997fq})
renormalization constants for $\alpha_s$ and the quark mass and {\em
 without} any necessity to renormalize the three auxiliary functions
themselves. The latter is
certainly possible but requires knowledge of three extra renormalization
constants, namely those for the ghost-ghost-gluon vertex, the gluon and
ghost wave functions. The corresponding calculation is certainly
possible but is {\em not} a small enterprise and the results for these
constants appeared only recently
\cite{Chetyrkin:2004mf,Czakon:2004bu}. The remarkable fact is that
the combination $T$ (defined in eq.~\re{inique_combination}) enters into
the renormalization of the  auxiliary functions as follows ($b$ 
is a known constant) 
\beq
\tilde\zeta_1^0 - \mbox{(known UV sub-divergencies)
 does not contain}  \ T
,
 \label{zeta10}
\eeq
\beq
\tilde\zeta_3^0 - \mbox{(known UV sub-divergencies)} = 1 +
b\,\alpha_s^3\,T  +\mbox{known terms} \label{zeta30:gh} {},
\eeq
and
\beq
\zeta_3^0 - \mbox{(known UV sub-divergencies)} =  1 - 2\,b\, \alpha_s^3\,T
+\mbox{known terms} \label{zeta30:gl} {}.
\eeq
We observe that 
\begin{itemize}
\item[(i)] the relation (\ref{extra_constraint}) can  be found in  the
process of renormalizing one of eqs.~(\ref{zeta30:gh} - \ref{zeta30:gl})
but {\em cannot} if one deals only with the physical quantity  $\zeta_g$.
\item[(ii)] our way of finding (\ref{extra_constraint})  is completely self
contained and does not require any knowledge of extra non-trivial
renormalization constants, while the same result obtained in
\cite{YS_MS} does depend on such knowledge. 
\item[(iii)] full agreement between relation (\ref{T91_0_N}) obtained 
in section 5 and eq.~(3.2) of  \cite{YS_MS} provides a
non-trivial check of the results in \cite{Chetyrkin:2004mf,Czakon:2004bu}.
\end{itemize}

Before concluding we want to comment on interesting questions
regarding the construction and the universality of the $\ep$-finite
basis.

The construction of the $\ep$-finite basis as discussed in
section \ref{sec:constr} requires that the reduction of the given set
of Feynman integrals to some basis of master integrals has already
been performed. For purely technical reasons this reduction procedure
is essentially more complicated and time consuming than finding the
$\ep$-finite basis. A significant part of the complexity arises from
the possible appearance of spurious poles since, as a consequence,  the exact
dependence on $d$ of all coefficient functions has to be kept through
the whole reduction procedure.

In a modified approach one could try to implement the idea of the
$\ep$-finite basis directly in the Laporta Algorithm, by e.g. properly
tuning specific weights in the algorithm. This would lead to a new
criterion instead of the commonly used one, which is based, roughly
speaking, on the number of lines, dots and the power of irreducible
numerators. This new criterion could, in addition,  depend on the
appearance of $(d-4)$ in the coefficients of the integrals, something
like ``never solve an equation with respect to an integral, which has
a factor of $(d-4)$ in its  coefficient''. As a result no spurious
pole would appear in any step, which would allow to expand the
coefficient functions in all equations around the point $d=4$
up to some fixed power in $\ep=(4-d)/2$ and would, in general,
lead to a significant gain in performance of the Laporta Algorithm.

On the other hand, it might happen that the IBP identities
``globally'' proportional to some positive power of $(d-4)$ play an
important role and that without their use no complete reduction is
possible. The problem is open and certainly deserves further
investigation.

The second interesting  problem is whether the $\ep$-finiteness of a basis is
preserved if one considers a different physical problem whose solution
is expressible in terms of the ``old'' master integrals. On general
grounds we can not exclude that for an ``old'' master, say, $T_i$ the
maximal power of the spurious pole would e.g. exceed the one displayed
on figure~\ref{fig:master}. In such a case the basis given in section
\ref{sec:constr.basis} does not have coefficient functions finite in the
limit $\ep \to 0$ and we would have to construct a new, more general
$\ep$-finite basis following the prescription of section~\ref{sec:constr}.  
However, we do think that such a situation is 
unlikely and that therefore the in section \ref{sec:constr.basis}
constructed $\ep$-finite basis is a universal one for all four-loop
problems reducible to the evaluation of QED like massive tadpoles. The
following observations support our point of view:
\begin{itemize}
\item
As stated in section 4 we have only used the reduction and integrals
appearing in the low energy expansion of the four-loop vacuum
polarization function (see \cite{Chetyrkin:2004fq}) for constructing
the $\ep$-finite basis. In fact, the evaluation of decoupling function
$\zeta_g$ for the QCD quark gluon coupling constant in four-loop
approximation requires the reduction of additional more complicated
integrals to masters. We have checked that the basis given in
figure~\ref{fig:epfb} remains $\ep$-finite for all these extra
integrals.
\item
If our hypothesis on the universality property of the constructed
$\ep$-finite basis is wrong it also implies that using the method
described in section \ref{subsect:T52} one should be able to find more
terms in the $\ep$-expansions of the ``old'' and of the $\ep$-finite
master integrals analytically. This extension of the equations
(\ref{tf52}-\ref{tf91},\ref{T71}-\ref{T72},\ref{extra_constraint})
without using more sophisticated techniques seems not very probable to 
us.
\end{itemize}

\subsection*{Conclusions}
As a conclusion we want to point out that the very idea of the
$\ep$-finite basis, namely the procedure of its systematic construction
as well as the application to finding {\em direct} analytic information
of $\ep${\em-finite} contributions to {\em  non-$\ep$-finite} master
integrals are all new. We believe that choosing the $\ep$-finite basis
should be especially useful and convenient in all cases when
numerical evaluation of master integrals is necessary.

Our results are also available in computer readable form
under the URL {\tt
http://www-ttp.physik.uni-karlsruhe.de/Progdata/ttp06/ttp06-02/}.

\subsection*{Acknowledgments}

The authors are grateful to R.V.~Harlander, J.H.~K\"uhn, Y.~Schr\"oder
and M.~Steinhauser for useful discussions. K.Ch. thanks M.~Kalmykov
for useful advice about the integral $T_{52}$.

The work was supported by the Deutsche Forschungsgemeinschaft in the
Sonderforschungsbereich/Transregio SFB/TR-9 ``Computational Particle
Physics''. 

The work of C.S. was also partially supported by MIUR
under contract 2001023713$\_$006.

\appendix
\section{Old  masters in terms of $\boldsymbol{\ep}$-finite basis
  \label{app:OldFromNew} }

\bea
\label{T52 }
&{}&T_{52} =
 \\ 
\nonumber
&+&
 \frac{2\spa (d-2)^2\spa (1298 - 953\spa d + 233\spa d^2 - 19\spa
   d^3)}{(d-3)\spa (3\spa d-10)\spa (3\spa d-8)\spa (73 - 34\spa d +
   4\spa d^2)} 
\,T_{41} 
 \\ 
\nonumber
&-&
 \frac{4\spa (d-5)\spa (d-4)\spa (4\spa d-17)}{(2\spa d-7)\spa (73 -
   34\spa d + 4\spa d^2)} 
\, T_{51} 
 \\ 
\nonumber
&+&
 \frac{16\spa (2\spa d-7)\spa (2\spa d-5)\spa (209 - 96\spa d + 11\spa
   d^2)}{3\spa (d-2)\spa (3\spa d-11)\spa (73 - 34\spa d + 4\spa d^2)} 
\, T_{53}  
 \\ 
\nonumber
&+&
 \frac{32\spa (d-4)\spa (d-3)^2\spa (2\spa d-7)\spa (149 - 69\spa d +
   8\spa d^2)}{(d-2)\spa (3\spa d-11)\spa (3\spa d-10)\spa (3\spa
   d-8)\spa (73 - 34\spa d + 4\spa d^2)} 
\, T_{63} 
 \\ 
\nonumber
&+&
 \frac{32\spa (d-5)\spa (d-4)^2\spa (d-3)}{(d-2)\spa (3\spa d-10)\spa
   (3\spa d-8)\spa (73 - 34\spa d + 4\spa d^2)} 
\, T^{f}_{52} 
 \\ 
\nonumber
&-&
 \frac{256\spa (d-5)\spa (d-4)^4}{3\spa (d-2)\spa (3\spa d-10)\spa
   (3\spa d-8)\spa (73 - 34\spa d + 4\spa d^2)} 
\,  T^{f}_{54} 
 \\ 
\nonumber
&-&
 \frac{16\spa (d-4)^3\spa (767 - 351\spa d + 40\spa d^2)}{3\spa
   (d-2)\spa (3\spa d-10)\spa (3\spa d-8)\spa (73 - 34\spa d + 4\spa
   d^2)} 
\, T^{f}_{61} 
 \\ 
\nonumber
&-&
 \frac{128\spa (d-5)^2\spa (d-4)^3}{(d-2)\spa (3\spa d-10)\spa (3\spa
   d-8)\spa (73 - 34\spa d + 4\spa d^2)} 
\, T^{f}_{62} 
 \\ 
\nonumber
&-&
 \frac{32\spa (d-4)^3\spa (109 - 51\spa d + 6\spa d^2)}{(d-2)\spa (3\spa
   d-10)\spa (3\spa d-8)\spa (73 - 34\spa d + 4\spa d^2)} 
\, T^{f}_{64} 
 \\ 
\nonumber
&-&
 \frac{48\spa (d-4)^3\spa (149 - 69\spa d + 8\spa d^2)}{(d-2)\spa (3\spa
   d-10)\spa (3\spa d-8)\spa (73 - 34\spa d + 4\spa d^2)} 
\, T^{f}_{71} 
 \\ 
\nonumber
&+&
 \frac{16\spa (d-4)^3\spa (673 - 297\spa d + 32\spa d^2)}{3\spa
   (d-2)\spa (3\spa d-10)\spa (3\spa d-8)\spa (73 - 34\spa d + 4\spa
   d^2)} 
\,  T^{f}_{72} 
 \\ 
\nonumber
&-&
 \frac{128\spa (d-5)\spa (d-4)^3\spa (3\spa d-13)}{3\spa (d-2)\spa
   (3\spa d-10)\spa (3\spa d-8)\spa (73 - 34\spa d + 4\spa d^2)} 
\, T_{91} \,,
\eea
\bea
\label{ T54 }
&{}& T_{54} =
 \\ 
\nonumber
&-&
 \frac{(d-2)^3\spa (21458 - 20723\spa d + 7488\spa d^2 - 1200\spa d^3 +
   72\spa d^4)}{2\spa (d-3)\spa (2\spa d-7)\spa (2\spa d-5)\spa (3\spa
   d-10)\spa (3\spa d-8)\spa (73 - 34\spa d + 4\spa d^2)} 
\,T_{41} 
 \\ 
\nonumber
&+&
 \frac{8\spa (d-4)^2\spa (d-2)\spa (25 - 22\spa d + 4\spa d^2)}{(2\spa
   d-7)^2\spa (2\spa d-5)\spa (73 - 34\spa d + 4\spa d^2)} 
\, T_{51} 
 \\ 
\nonumber
&-&
 \frac{32\spa (d-4)\spa (10\spa d-33)}{3\spa (3\spa d-11)\spa (73 -
   34\spa d + 4\spa d^2)} 
\, T_{53} 
 \\ 
\nonumber
&-&
 \frac{256\spa (d-4)^2\spa (d-3)^3}{(2\spa d-5)\spa (3\spa d-11)\spa
   (3\spa d-10)\spa (3\spa d-8)\spa (73 - 34\spa d + 4\spa d^2)} 
\, T_{63} 
 \\ 
\nonumber
&+&
 \frac{256\spa (d-5)\spa (d-4)^3\spa (d-3)}{(2\spa d-7)\spa (2\spa
   d-5)\spa (3\spa d-10)\spa (3\spa d-8)\spa (73 - 34\spa d + 4\spa
   d^2)} 
\, T^{f}_{52} 
 \\ 
\nonumber
&+&
 \frac{512\spa (d-4)^4}{3\spa (2\spa d-5)\spa (3\spa d-10)\spa (3\spa
   d-8)\spa (73 - 34\spa d + 4\spa d^2)} 
\,  T^{f}_{54} 
 \\ 
\nonumber
&+&
 \frac{128\spa (d-4)^4\spa (11\spa d-37)}{3\spa (2\spa d-7)\spa (2\spa
   d-5)\spa (3\spa d-10)\spa (3\spa d-8)\spa (73 - 34\spa d + 4\spa
   d^2)} \, T^{f}_{61} 
 \\ 
\nonumber
&+&
 \frac{256\spa (d-5)\spa (d-4)^3}{(2\spa d-5)\spa (3\spa d-10)\spa
   (3\spa d-8)\spa (73 - 34\spa d + 4\spa d^2)} \, T^{f}_{62} 
 \\ 
\nonumber
&+& \frac{128\spa (d-4)^4}{(2\spa d-7)\spa (2\spa d-5)\spa (3\spa
  d-10)\spa (3\spa d-8)\spa (73 - 34\spa d + 4\spa d^2)} \, T^{f}_{64}
 \\ 
\nonumber
&+& \frac{384\spa (d-4)^4\spa (d-3)}{(2\spa d-7)\spa (2\spa d-5)\spa
  (3\spa d-10)\spa (3\spa d-8)\spa (73 - 34\spa d + 4\spa d^2)} \,
T^{f}_{71}
 \\ 
\nonumber
&-& \frac{128\spa (d-4)^4\spa (25\spa d-89)}{3\spa (2\spa d-7)\spa
  (2\spa d-5)\spa (3\spa d-10)\spa (3\spa d-8)\spa (73 - 34\spa d +
  4\spa d^2)} \,  T^{f}_{72} 
 \\ 
\nonumber
&+& \frac{256\spa (d-4)^3\spa (3\spa d-13)}{3\spa (2\spa d-5)\spa (3\spa
  d-10)\spa (3\spa d-8)\spa (73 - 34\spa d + 4\spa d^2)} \, T_{91} \,,
\eea
\bea
\label{ T64 }
&{}& T_{64} =
 \\ 
\nonumber
&+& \frac{3\spa (d-4)\spa (d-2)\spa (3\spa d-8)\spa (82 - 48\spa d +
  7\spa d^2)}{64\spa (d-3)^3\spa (2\spa d-7)^2} \, T_{51} 
 \\ 
\nonumber
&-& \frac{(2\spa d-5)\spa (3\spa d-10)^2\spa (3\spa d-8)}{4\spa
  (d-4)\spa (d-3)^3} \, T_{53} 
 \\ 
\nonumber
&-&
 \frac{3\spa (3\spa d-10)}{4\spa (d-3)}
\, T_{63} 
 \\ 
\nonumber
&+&
 \frac{9\spa (d-4)^2\spa (3\spa d-11)\spa (3\spa d-10)}{8\spa
   (d-3)^3\spa (2\spa d-7)} \, T^{f}_{71} \,, 
\eea
\bea
\label{ T61 }
&{}& T_{61} =
 \\ 
\nonumber
&+&
 \frac{3\spa (d-2)^3\spa (11278 - 11551\spa d + 4411\spa d^2 - 745\spa
   d^3 + 47\spa d^4)}{16\spa (d-3)^3\spa (2\spa d-7)\spa (3\spa
   d-10)\spa (73 - 34\spa d + 4\spa d^2)} 
\,T_{41} 
 \\ 
\nonumber
&+&
 \frac{3\spa (d-5)\spa (d-4)\spa (d-2)\spa (3\spa d-8)\spa (4\spa
   d-17)\spa (5\spa d-16)}{8\spa (d-3)^2\spa (2\spa d-7)^2\spa (73 -
   34\spa d + 4\spa d^2)} 
\, T_{51} 
 \\ 
\nonumber
&-&
 \frac{(2\spa d-5)\spa (3\spa d-8)\spa (21406 - 22007\spa d + 8444\spa
   d^2 - 1434\spa d^3 + 91\spa d^4)}{2\spa (d-4)\spa (d-3)^2\spa (3\spa
   d-11)\spa (73 - 34\spa d + 4\spa d^2)} 
\, T_{53} 
 \\ 
\nonumber
&-&
 \frac{3\spa (17566 - 18119\spa d + 6980\spa d^2 - 1191\spa d^3 + 76\spa
   d^4)}{(3\spa d-11)\spa (3\spa d-10)\spa (73 - 34\spa d + 4\spa d^2)} 
\, T_{63} 
 \\ 
\nonumber
&-&
 \frac{3\spa (d-5)\spa (d-4)^2\spa (5\spa d-16)}{(d-3)\spa (2\spa
   d-7)\spa (3\spa d-10)\spa (73 - 34\spa d + 4\spa d^2)} 
\, T^{f}_{52} 
 \\ 
\nonumber
&+&
 \frac{8\spa (d-5)\spa (d-4)^4\spa (5\spa d-16)}{(d-3)^2\spa (2\spa
   d-7)\spa (3\spa d-10)\spa (73 - 34\spa d + 4\spa d^2)} 
\,  T^{f}_{54} 
 \\ 
\nonumber
&+&
 \frac{(d-4)^2\spa (73178 - 75093\spa d + 28748\spa d^2 - 4869\spa d^3 +
   308\spa d^4)}{2\spa (d-3)^2\spa (2\spa d-7)\spa (3\spa d-10)\spa (73
   - 34\spa d + 4\spa d^2)} 
\, T^{f}_{61} 
 \\ 
\nonumber
&+&
 \frac{12\spa (d-5)^2\spa (d-4)^3\spa (5\spa d-16)}{(d-3)^2\spa (2\spa
   d-7)\spa (3\spa d-10)\spa (73 - 34\spa d + 4\spa d^2)} 
\, T^{f}_{62} 
 \\ 
\nonumber
&+&
 \frac{3\spa (d-4)^2\spa (15006 - 15527\spa d + 6004\spa d^2 - 1029\spa
   d^3 + 66\spa d^4)}{(d-3)^2\spa (2\spa d-7)\spa (3\spa d-10)\spa (73 -
   34\spa d + 4\spa d^2)} 
\, T^{f}_{64} 
 \\ 
\nonumber
&+&
 \frac{9\spa (d-4)^2\spa (17566 - 18119\spa d + 6980\spa d^2 - 1191\spa
   d^3 + 76\spa d^4)}{2\spa (d-3)^2\spa (2\spa d-7)\spa (3\spa d-10)\spa
   (73 - 34\spa d + 4\spa d^2)} 
\, T^{f}_{71} 
 \\ 
\nonumber
&-&
 \frac{(d-4)^2\spa (18982 - 18219\spa d + 6388\spa d^2 - 963\spa d^3 +
   52\spa d^4)}{2\spa (d-3)^2\spa (2\spa d-7)\spa (3\spa d-10)\spa (73 -
   34\spa d + 4\spa d^2)} 
\,  T^{f}_{72} 
 \\ 
\nonumber
&+&
 \frac{4\spa (d-5)\spa (d-4)^3\spa (3\spa d-13)\spa (5\spa
   d-16)}{(d-3)^2\spa (2\spa d-7)\spa (3\spa d-10)\spa (73 - 34\spa d +
   4\spa d^2)} 
\, T_{91} \,,
\eea
\bea
\label{ T62 }
&{}& T_{62} =
 \\ 
\nonumber
&-&
 \frac{(d-4)^2\spa (d-2)^3\spa (d-1)}{4\spa (d-3)^3\spa (2\spa d-7)\spa
   (73 - 34\spa d + 4\spa d^2)} 
\,T_{41} 
 \\ 
\nonumber
&+&
 \frac{(d-2)\spa (3\spa d-8)\spa (499 - 409\spa d + 111\spa d^2 - 10\spa
   d^3)}{4\spa (d-3)^2\spa (2\spa d-7)\spa (73 - 34\spa d + 4\spa d^2)} 
\, T_{51} 
 \\ 
\nonumber
&-&
 \frac{2\spa (2\spa d-5)\spa (3\spa d-10)\spa (3\spa d-8)\spa (63 -
   28\spa d + 3\spa d^2)}{3\spa (d-3)^2\spa (3\spa d-11)\spa (73 -
   34\spa d + 4\spa d^2)} 
\, T_{53} 
 \\ 
\nonumber
&-&
 \frac{4\spa (d-4)^2\spa (4\spa d-19)}{(3\spa d-11)\spa (73 - 34\spa d +
   4\spa d^2)} 
\, T_{63} 
 \\ 
\nonumber
&-&
 \frac{4\spa (d-5)\spa (d-4)^2}{(d-3)\spa (2\spa d-7)\spa (73 - 34\spa d
   + 4\spa d^2)} 
\, T^{f}_{52} 
 \\ 
\nonumber
&-&
 \frac{16\spa (d-4)^3\spa (33 - 16\spa d + 2\spa d^2)}{3\spa (d-3)^2\spa
   (2\spa d-7)\spa (73 - 34\spa d + 4\spa d^2)} 
\,  T^{f}_{54} 
 \\ 
\nonumber
&-&
 \frac{2\spa (d-4)^3\spa (255 - 125\spa d + 16\spa d^2)}{3\spa
   (d-3)^2\spa (2\spa d-7)\spa (73 - 34\spa d + 4\spa d^2)} 
\, T^{f}_{61} 
 \\ 
\nonumber
&-&
 \frac{4\spa (d-5)\spa (d-4)^2}{(d-3)^2\spa (73 - 34\spa d + 4\spa d^2)}
\, T^{f}_{62} 
 \\ 
\nonumber
&+&
 \frac{4\spa (d-4)^4\spa (2\spa d-9)}{(d-3)^2\spa (2\spa d-7)\spa (73 -
   34\spa d + 4\spa d^2)} 
\, T^{f}_{64} 
 \\ 
\nonumber
&+&
 \frac{6\spa (d-4)^4\spa (4\spa d-19)}{(d-3)^2\spa (2\spa d-7)\spa (73 -
   34\spa d + 4\spa d^2)} 
\, T^{f}_{71} 
 \\ 
\nonumber
&+&
 \frac{2\spa (d-4)^3\spa (276 - 145\spa d + 20\spa d^2)}{3\spa
   (d-3)^2\spa (2\spa d-7)\spa (73 - 34\spa d + 4\spa d^2)} 
\,  T^{f}_{72} 
 \\ 
\nonumber
&-&
 \frac{8\spa (d-4)^2\spa (3\spa d-13)\spa (33 - 16\spa d + 2\spa
   d^2)}{3\spa (d-3)^2\spa (2\spa d-7)\spa (73 - 34\spa d + 4\spa d^2)} 
\, T_{91} \,,
\eea
\bea
\label{ T72 }
&{}& T_{72} =
 \\ 
\nonumber
&-&
 \frac{(d-2)^3\spa (1789 - 2111\spa d + 912\spa d^2 - 172\spa d^3 +
   12\spa d^4)}{8\spa (d-3)^2\spa (2\spa d-7)\spa (3\spa d-10)^2\spa (73
   - 34\spa d + 4\spa d^2)} 
\,T_{41} 
 \\ 
\nonumber
&-&
 \frac{(d-4)\spa (d-2)\spa (3\spa d-8)\spa (914 - 525\spa d + 90\spa d^2
   - 4\spa d^3)}{16\spa (d-3)\spa (2\spa d-7)^2\spa (3\spa d-10)\spa (73
   - 34\spa d + 4\spa d^2)} 
\, T_{51} 
 \\ 
\nonumber
&+&
 \frac{(2\spa d-5)\spa (3\spa d-8)\spa (9772 - 7470\spa d + 1165\spa d^2
   + 399\spa d^3 - 142\spa d^4 + 12\spa d^5)}{6\spa (d-4)^2\spa
   (d-3)\spa (3\spa d-11)\spa (3\spa d-10)\spa (73 - 34\spa d + 4\spa
   d^2)} 
\, T_{53} 
 \\ 
\nonumber
&+&
 \frac{(d-3)^2\spa (12628 - 13328\spa d + 5255\spa d^2 - 918\spa d^3 +
   60\spa d^4)}{(d-4)\spa (3\spa d-11)\spa (3\spa d-10)^2\spa (73 -
   34\spa d + 4\spa d^2)} 
\, T_{63} 
 \\ 
\nonumber
&-&
 \frac{8\spa (d-5)\spa (d-4)^3}{(2\spa d-7)\spa (3\spa d-10)^2\spa (73 -
   34\spa d + 4\spa d^2)} 
\, T^{f}_{52} 
 \\ 
\nonumber
&-&
 \frac{16\spa (d-4)^4}{3\spa (d-3)\spa (3\spa d-10)^2\spa (73 - 34\spa d
   + 4\spa d^2)} 
\,  T^{f}_{54} 
 \\ 
\nonumber
&-&
 \frac{4\spa (d-4)^4\spa (11\spa d-37)}{3\spa (d-3)\spa (2\spa d-7)\spa
   (3\spa d-10)^2\spa (73 - 34\spa d + 4\spa d^2)} 
\, T^{f}_{61} 
 \\ 
\nonumber
&-&
 \frac{8\spa (d-5)\spa (d-4)^3}{(d-3)\spa (3\spa d-10)^2\spa (73 -
   34\spa d + 4\spa d^2)} 
\, T^{f}_{62} 
 \\ 
\nonumber
&+&
 \frac{2\spa (d-4)\spa (35898 - 47927\spa d + 25554\spa d^2 - 6802\spa
   d^3 + 904\spa d^4 - 48\spa d^5)}{(d-3)\spa (2\spa d-7)\spa (3\spa
   d-10)^2\spa (73 - 34\spa d + 4\spa d^2)} 
\, T^{f}_{64} 
 \\ 
\nonumber
&-&
 \frac{3\spa (d-4)\spa (12628 - 13328\spa d + 5255\spa d^2 - 918\spa d^3
   + 60\spa d^4)}{2\spa (2\spa d-7)\spa (3\spa d-10)^2\spa (73 - 34\spa
   d + 4\spa d^2)} 
\, T^{f}_{71} 
 \\ 
\nonumber
&+&
 \frac{4\spa (d-4)^4\spa (25\spa d-89)}{3\spa (d-3)\spa (2\spa d-7)\spa
   (3\spa d-10)^2\spa (73 - 34\spa d + 4\spa d^2)} 
\,  T^{f}_{72} 
 \\ 
\nonumber
&-&
 \frac{8\spa (d-4)^3\spa (3\spa d-13)}{3\spa (d-3)\spa (3\spa
   d-10)^2\spa (73 - 34\spa d + 4\spa d^2)} 
\, T_{91} \,,
\eea
\bea
\label{ T71 }
&{}& T_{71} =
 \\ 
\nonumber
&-&
 \frac{(d-2)^3\spa (5925 - 6293\spa d + 2499\spa d^2 - 440\spa d^3 +
   29\spa d^4)}{4\spa (d-3)^2\spa (2\spa d-7)\spa (3\spa d-10)^2\spa (73
   - 34\spa d + 4\spa d^2)} 
\,T_{41} 
 \\ 
\nonumber
&-&
 \frac{(d-2)\spa (3\spa d-8)\spa (9053 - 9587\spa d + 3799\spa d^2 -
   668\spa d^3 + 44\spa d^4)}{4\spa (d-3)\spa (2\spa d-7)^2\spa (3\spa
   d-10)\spa (73 - 34\spa d + 4\spa d^2)} 
\, T_{51} 
 \\ 
\nonumber
&-&
 \frac{2\spa (2\spa d-5)\spa (3\spa d-8)\spa (26546 - 35605\spa d +
   19114\spa d^2 - 5132\spa d^3 + 689\spa d^4 - 37\spa d^5)}{3\spa
   (d-4)^2\spa (d-3)\spa (3\spa d-11)\spa (3\spa d-10)\spa (73 - 34\spa
   d + 4\spa d^2)} 
\, T_{53} 
 \\ 
\nonumber
&-&
 \frac{4\spa (d-3)\spa (19090 - 25909\spa d + 14064\spa d^2 - 3815\spa
   d^3 + 517\spa d^4 - 28\spa d^5)}{(d-4)\spa (3\spa d-11)\spa (3\spa
   d-10)^2\spa (73 - 34\spa d + 4\spa d^2)} 
\, T_{63} 
 \\ 
\nonumber
&-&
 \frac{4\spa (d-5)\spa (d-4)^2\spa (5\spa d-19)}{(2\spa d-7)\spa (3\spa
   d-10)^2\spa (73 - 34\spa d + 4\spa d^2)} 
\, T^{f}_{52} 
 \\ 
\nonumber
&+&
 \frac{16\spa (d-4)^3\spa (554 - 435\spa d + 114\spa d^2 - 10\spa
   d^3)}{3\spa (d-3)\spa (2\spa d-7)\spa (3\spa d-10)^2\spa (73 - 34\spa
   d + 4\spa d^2)} 
\,  T^{f}_{54} 
 \\ 
\nonumber
&+&
 \frac{2\spa (d-4)^3\spa (7327 - 5682\spa d + 1473\spa d^2 - 128\spa
   d^3)}{3\spa (d-3)\spa (2\spa d-7)\spa (3\spa d-10)^2\spa (73 - 34\spa
   d + 4\spa d^2)} 
\, T^{f}_{61} 
 \\ 
\nonumber
&-&
 \frac{4\spa (d-5)\spa (d-4)^2\spa (54 - 29\spa d + 4\spa
   d^2)}{(d-3)\spa (3\spa d-10)^2\spa (73 - 34\spa d + 4\spa d^2)} 
\, T^{f}_{62} 
 \\ 
\nonumber
&+&
 \frac{4\spa (d-4)\spa (21906 - 29301\spa d + 15680\spa d^2 - 4195\spa
   d^3 + 561\spa d^4 - 30\spa d^5)}{(d-3)\spa (2\spa d-7)\spa (3\spa
   d-10)^2\spa (73 - 34\spa d + 4\spa d^2)} 
\, T^{f}_{64} 
 \\ 
\nonumber
&+&
 \frac{6\spa (d-4)\spa (19090 - 25909\spa d + 14064\spa d^2 - 3815\spa
   d^3 + 517\spa d^4 - 28\spa d^5)}{(d-3)\spa (2\spa d-7)\spa (3\spa
   d-10)^2\spa (73 - 34\spa d + 4\spa d^2)} 
\, T^{f}_{71} 
 \\ 
\nonumber
&-&
 \frac{2\spa (d-4)\spa (50762 - 63841\spa d + 32144\spa d^2 - 8107\spa
   d^3 + 1025\spa d^4 - 52\spa d^5)}{3\spa (d-3)\spa (2\spa d-7)\spa
   (3\spa d-10)^2\spa (73 - 34\spa d + 4\spa d^2)} 
\,  T^{f}_{72} 
 \\ 
\nonumber
&+&
 \frac{8\spa (d-4)^2\spa (3\spa d-13)\spa (554 - 435\spa d + 114\spa d^2
   - 10\spa d^3)}{3\spa (d-3)\spa (2\spa d-7)\spa (3\spa d-10)^2\spa (73
   - 34\spa d + 4\spa d^2)} 
\, T_{91} \,.
\eea

\section{Proceeding in case of a ``divergent'' cut\label{app:diaB}}

In this section we show an alternative approach of calculating diagram
\diaA. If we perform the cut of the diagram in the following way
\begin{equation}
  \label{eq:alt.self.diaA}
  \cutdia{\diaA}{\intmom} =
  \parbox{4cm}{\includegraphics[width=4cm]{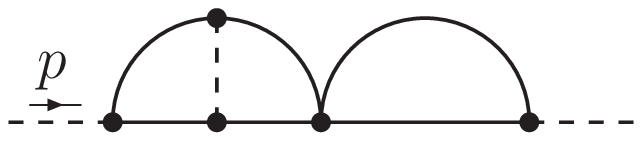}}
\end{equation}
one obtains a divergent self energy function
$\cutdia{\diaA}{\intmom}$. This results from the fact that the diagram
in equation \re{eq:alt.self.diaA} consists of
\begin{equation}
  \label{eq:alt.diaA.mult}
\parbox{4cm}{\includegraphics[width=4cm]{t52fcut2.eps}} = 
\underbrace{\parbox{2.7cm}{\includegraphics[width=2.7cm%
]{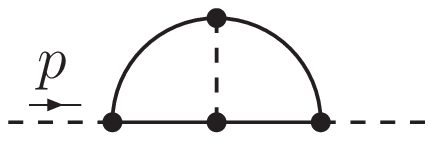}}}_{=\text{finite}}
\times 
\underbrace{\parbox{2.7cm}{\includegraphics[width=2.7cm%
]{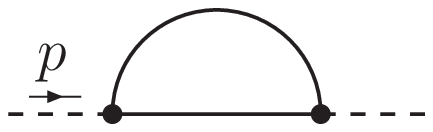}}}_{= \frac 1{\ep} + \text{finite}}
\end{equation}
a finite two-loop multiplied by a divergent one-loop diagram. If one
identifies such a sub-divergency one is able to construct the combination
\begin{equation}
    \cutdia{\diaA}{\intmom} =
\underbrace{\parbox{2.7cm}{\includegraphics[width=2.7cm%
]{t52fcut2part1.eps}}
  \times \left(
    \parbox{2.7cm}{\includegraphics[width=2.7cm]{t52fcut2part2.eps}} -
    \frac 1{\ep} \right)}_{\equiv \cutdiafinite{\diaA}{\intmom}
  =\text{finite}}  +\, \frac 1{\ep}
\parbox{2.7cm}{\includegraphics[width=2.7cm]{t52fcut2part1.eps}}
\end{equation}
of a finite self energy function $\cutdiafinite{\diaA}{\intmom}$ and a
diagram of lower loop order which can be calculated analytically.
In general one might have to treat several such sub-divergencies in the
similar way. In a next step $\cutdiafinite{\diaA}{\intmom}$
\begin{equation}
  \cutdiafinite{\diaA}{\intmom} =
  \parbox{4cm}{\includegraphics[width=4cm]{t52fcut2.eps}} -
\frac 1{\ep}
\parbox{2.7cm}{\includegraphics[width=2.7cm]{t52fcut2part1.eps}}
\end{equation}
is evaluated to obtain ${\tilde{T}^f}_{52 \, [{\paden}/{\padem}]}$
according to the proceeding described in section \ref{sec:pade}. Thus we
are again able to extract the pole structure of ${\tilde{T}^f}_{52 \,
  [{\paden}/{\padem}]}$ analytically and the constant part of the
$\ep$-expansion numerically. Finally one obtains the expansion for the
diagram \diaA 
\begin{equation}
\diaAeq_{[{\paden}/{\padem}]} = {\tilde{T}^f}_{52 \,
  [{\paden}/{\padem}]} + \frac 1{\ep}
\parbox{1.7cm}{\includegraphics[width=1.7cm]{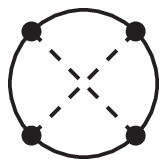}}
\end{equation}
by adding the analytical result of the three-loop tadpole diagram. We
find full agreement between this approach and the results discussed in
section \ref{sec:pade}.

\end{document}